\documentstyle[12pt]{article}
\begin{document}

\begin{titlepage}
\flushright{\bf IPGAS-HEP-TH/30/96\\
            July, 1996     }

\vspace{3cm}

\begin{center}
{\bf ON THE POSSIBLE SOLUTION OF THE DOUBLET-TRIPLET SPLITTING \\
PROBLEM IN EXTENDED $SU(N\geq 8)$ SUSY GUTS}

\vspace{1cm}

A.Kobakhidze \footnote{E-mail:~ yoko@physics.iberiapac.ge}

\vspace{0.5cm}

Institute of Physics, Tamarashvili str.6, GE-380077 Tbilisi, Georgia
\end{center}
\begin{abstract}
The generalization of the "custodial symmetry" mechanism which leads
to the natural (without fine tuning) explanation of the
doublet-triplet hierarchy is suggested in the frames of extended
$SU(N\geq 8)$ SUSY GUTs. It is shown also that such type of SUSY
GUT can predict the value of strong gauge coupling constant
$\alpha _S(M_Z)$ consistent with present experimental data.
\end{abstract}
\end{titlepage}

\section{Introduction}

From the time of birth of Grand Unified Theories (GUTs) the gauge hierarchy
problem remains as one of the challenging task of elementary particle
physics. For the solution of this problem it is necessary to answer the
following two questions:
\begin{quote}
1.~Why the electroweak scale ($M_W\sim 10^2GeV$) is so small in comparison
with Grand ($M_G\sim 10^{16}GeV$) or Plank ($M_{Pl}.\sim 10^{19}GeV$)
scales? \\ 2.~Why is the electroweak scale stable under the radiative
corrections?
\end{quote}

The second problem is solved by supersymmetry (SUSY), which ensures
stability of any mass scale due to the "nonrenormalization" theorem [1].
This is the main motivation for introduction of the low-energy ($
M_{SUSY}\sim 1 TeV$) broken SUSY. Moreover, some times ago, LEP precision
measurements of the Standard Model (SM) gauge couplings $\alpha _S,
\alpha _W $ and $\alpha _Y$ at Z-pick gives an evidence of the
perfect unification of the running gauge couplings in the minimal
SUSY extension of SU(5) GUT and finally ruled out non-SUSY SU(5)
model [2]. This fact was considered by many physicists as a hint on
the existence of low-energy broken SUSY in nature\footnote{However,
calculations of the running gauge couplings including detailed
structure of SUSY and GUT thresholds suggested in [3] showed that the
minimal version of SUSY SU(5) predicts the values of $\alpha _S(M_Z)$
which are inconsistent with those extracted from low-energy
experiments. Several attempts to decrease the predicted value of
$\alpha _S(M_Z)$ have been made in the frames of extended SUSY GUTs
[4] or string type unification models [5]. On the other hand, there
was known several examples of successful unification in non-SUSY
extension of the minimal SU(5) [6].It was shown also that some of
these models predicts the correct $b$-quark mass from the requirement
of the exact $b-\tau$ unification as well [7].}.

However, the straightforward SUSY extension of the ordinary GUTs can
not give an answer to the first question. In order to have
phenomenologically desirable picture one must provide light ($\sim
M_W$) masses for electroweak doublets and at the same time keep their
colored triplet partners superheavy ($\sim M_G$). In the opposite
case, colored triplets lead to unacceptable fast proton decay, except
the case when they do not couple to light generation fermions [8],
and heavy doublets give an unacceptable large electroweak symmetry
breaking scale. This is the famous doublet-triplet splitting problem.
The simplest solution of this problem is the appropriate fine tuning
of the tree level superpotential parameters (so called "technical"
solution) [9], but it seemed to be extremely unnatural.

Several attempts have been made to explain the doublet-triplet
hierarchy problem in a natural (without fine tuning) way, such as:
"sliding singlet" mechanism [10], which is either unstable under the
SUSY breaking [11] or to be implemented in extended GUTs requires
complicated Higgs sector [12]; "missing doublet" mechanism also
requires Higgs superfields in a very complicate representations ($75
+ 50 + \overline {50}$ in the SU(5) case) [13]; "missing VEV"
mechanism [14]; GIFT mechanism, where Higgs doublets are identified
with pseudogoldstone particles of the spontaneously broken global
symmetry [15]; "custodial symmetry" mechanism [16]; "missing doublet
representation" mechanism, which explore combined features of some
above mentioned mechanisms [17].

In the present paper I suggest the generalization of the ''custodial
symmetry'' mechanism for the explanation of the doublet-triplet
hierarchy proposed some times ago by G.Dvali [16]. The main idea of
this mechanism is the following: the Higgs doublet will be
automatically light if it is related by a certain ''custodial''
symmetry to another doublet which after the GUT symmetry breaking
becomes an eaten up goldstone particle. It is clear, that the
necessary condition for the realization of this mechanism is that the
GUT symmetry group being broken down to the $G_{SM}\equiv
SU(3)_C\otimes SU(2)_W\otimes U(1)_Y$ must produce an unphysical
Higgs with quantum numbers of electriweak doublet. The minimal group
satisfying this condition is SU(6). In [16] had been proved the
following theorem:

\begin{quote} Imagine an $SU(6)\otimes SU(N)_{c}$ invariant theory
with arbitrary renormalizable superpotential involving Higgs
superfields: $\Sigma ^i_j \sim (35.1), H_{i\alpha}\sim (6.N),
\overline {H}^{i\alpha}\sim (\overline 6.\overline N)$, where
$SU(N)_{c}$ is an additional "custodial" symmetry (here latin indices
stand for SU(6) group and those of greek for $SU(N)_{c}$ one). If
there exists a supersymmetric $G_{SM}$ invariant minimum in which $
SU(3)_W$ is predominantly (up to the admixtures of the order of
$M_W$) broken by vacuum expectation value (VEV) of the pair
$H_{i\alpha}+\overline {H}^{i\alpha}$, then $N-1$ pair of electroweak
doublets are necessarily light with masses $\sim M_W$.  \end{quote}

So, in order to have one pair of light doublets in addition to the
local SU(6) one needs $SU(2)_{c}$ symmetry which relates light Higgs
doublets with eaten up goldstone eigenstates.

However, the general analysis of the $SU(6)\otimes SU(N)_c$ invariant
superpotential shows that in the ''Grand desert'' region at the scale
of the order of $M_I\sim (M_WM_G)^{1/2}$ necessarily exists
intermediate $G_I\equiv SU(3)_C\otimes SU(3)_W\otimes U(1)$ symmetry
structure which spoils the standard unification picture.

As it will be shown below it is possible to generalize this mechanism
by embedding the local "custodial symmetry" into higher GUTs and at
the same time to resolve the unification problem arising in the
simplest version of the model [16].

\section{$SU(6)\otimes SU(2)_c$ model}

Before getting the generalization of the ''custodial symmetry''
mechanism let us briefly consider the original proposal [16]. As it
was described in the previous section, the minimal model based on
$SU(6)\otimes SU(2)_c$ symmetry group, where $SU(2)_c$ is global
''custodial symmetry'' under which one pair of SU(6) fundamental
(antifundamental) Higgs superfields transforms as doublet
(antidoublet): $H_{i\alpha }+\overline{H}^{i\alpha }$. In addition
one has SU(6) adjoint Higgs superfields $\Sigma _j^i$. Then the most
general renormalizable (cubic) superpotential involving these
superfields \begin{equation} \label{1}W(\Sigma ,H,\overline{H})=\frac
12M_\Sigma Tr\Sigma ^2+\frac 13\sigma Tr\Sigma ^3+M_HH_{i\alpha
}\overline{H}^{i\alpha }+hH_{i\alpha }\Sigma
_j^i\overline{H}^{j\alpha } \end{equation} among the other SUSY
degenerate vacua posses the following one:  \begin{eqnarray}{l} <
\Sigma > = diag\biggl [ ~1~ 1~ 1~ -1~ -1~ -1~ \biggr ] \cdot
\frac{M_H}{h} + diag\biggl [ ~2 ~2 ~2 ~-3 ~-3 ~0 \biggr ] \cdot
\frac{M_{\Sigma}}{\sigma} , \nonumber  \\ \nonumber
              \\ < H > = \left[ \begin{array}{cc} 0 & 0 \\ 0 & 0 \\ 0
& 0 \\ 0 & 0 \\ 0 & 0 \\ V & 0 \end{array} \right] ,~~~< \overline
{H} > = \left[ \begin{array}{cccccc} 0 & 0 & 0 & 0 & 0 & V \\ 0 & 0 &
0 & 0 & 0 & 0 \end{array} \right] ,~~ where~ V = \biggl
[\frac{6M_{\Sigma}}{h}\biggl (\frac{M_{\Sigma}}{\sigma} +
\frac{M_H}{h}\biggr )\biggr ]^{1/2} \label{2} \end{eqnarray} Suppose
that $M_\Sigma$ is of the order of $M_W$ and $M_H\sim M_G$. In this
case SU(6) local symmetry is broken through $S(3)_C\otimes
SU(3)_W\otimes U(1)$ channel and $SU(3)_W$ is predominantly broken by
the VEV of $H_{i1} + \overline {H}^{i1}$, so the condition of the
general theorem is holded. It is easy to verify that pair of Higgs
doublets from $H_{i2} + \overline {H}^{i2}$ has mass
$\frac{-3M_{\Sigma}}{h}\sim M_W$ while colored triplets are
superheavy ($\sim M_H\sim M_G$). However one can see from Fig.1 that
hierarchical SU(6) breaking \begin{equation}
\label{3}SU(6)\stackrel{M_G}{\longrightarrow}SU(3)_C\otimes
SU(3)_W\otimes U(1)\stackrel{\sim
(M_GM_W)^{1/2}}{\longrightarrow}G_{SM} \end{equation} spoils the
standard unification picture due to the influence of new gauge
interactions on the running of gauge couplings above the intermediate
symmetry scale. So, while achieving natural doublet-triplet splitting
in a simple and elegant way at the same time we loose the gauge
coupling unification. Thus model is inconsistent.
\begin{figure}
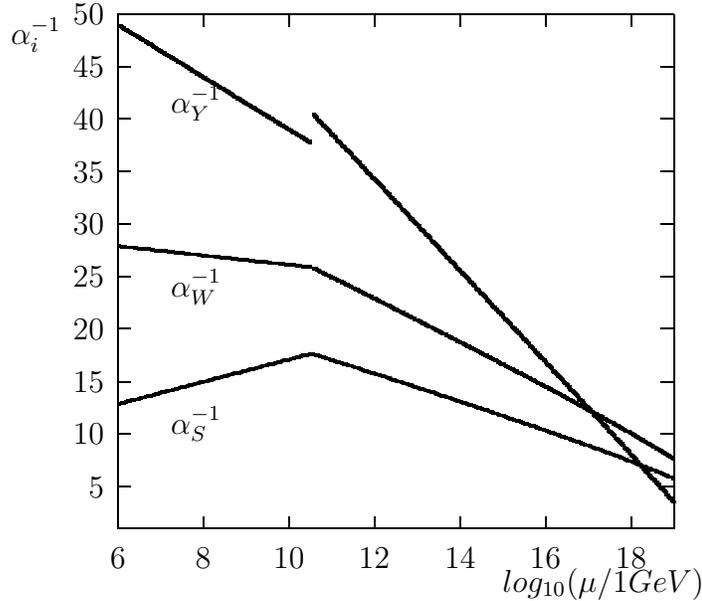

\begin{center}
\setlength{\unitlength}{0.240900pt}
\ifx\plotpoint\undefined\newsavebox{\plotpoint}\fi
\sbox{\plotpoint}{\rule[-0.175pt]{0.350pt}{0.350pt}}%

\end{center}
\caption{\small {The running of gauge couplings in $SU(6)\otimes
SU(2)_c$ model of ref.[16] for $M_{SUSY}=M_Z$, $\alpha
_S(M_Z)=0.118$, $sin^2\theta _W(M_Z)=0.2312$ and $\alpha
^{-1}_{EM}=127.9$}.} \end{figure}

\section{SU(8) generalization}

Now let us consider the generalization of the model discussed above.
Let me suppose that instead of the global $SU(2)_c$, dynamical origin
of which is very hard to understand, this symmetry is global and
moreover, it is embedded together with SU(6) into the simple gauge
group. The minimal unitary group of such type is SU(8). The minimal
generalization of the Higgs superfield content responsible at the
same time to the complete SU(8) breaking down to the $G_{SM}$ is the
following (latin indices now are those of SU(8)):  \begin{equation}
\label{4}\Sigma ^i_j\sim 63 ,~~H_{[ij]}\sim 28 ,~~\overline
{H}^{[ij]}\sim \overline 28 ,~~\Phi _{i1}\sim 8 ,~~\Phi _{i2}\sim 8
,~~\overline {\Phi}^{i}_1\sim \bar 8 ,~~\overline {\Phi}^{i}_2\sim
\overline 8 \end{equation} In order to eliminate from the
superpotential undesirable couplings such as $\overline H\Phi _1\Phi
_2 + H\overline {\Phi}_1\overline {\Phi}_2$ one can require
additional discrete symmetry under which $\Phi _2$ and $\overline
{\Phi}_2$ only change sign \footnote{ This discrete symmetry is
anomaly free and could be considered as a discrete gauge symmetry.
Note that introduction of $\Phi _2$ and $\overline {\Phi}_2$
superfields and subsequently of the discrete symmetry is not
necessary for the solution of the doublet-triplet splitting problem
and complete SU(8) breaking down to the $G_{SM}$, but as will be seen
below they play crucial role in the unification of gauge couplings}.
Then the general superpotential looks like:  \begin{equation}
\label{5}\frac{\sigma}{3}Tr\Sigma ^3 + M_HH_{[ij]}\overline
{H}^{[ij]} + hH_{[ij]}\Sigma ^i_k\overline {H}^{[ki]} + a\Phi
_{i1}\Sigma ^i_j \overline {\Phi}^j_1 + M_{\Phi}\Phi _{i2}\overline
{\Phi}^i_2 + b\Phi _{i2}\Sigma ^i_j\overline {\Phi}^j_2 ~,
\end{equation} where I have omitted the mass terms of $\Sigma ^2$ and
$\Phi _2\overline {\Phi}_2$ assuming that they are small ($\sim M_W$)
in comparison with $M_H$ and $M_\Phi$ and they can easily be
reproduced in the scalar superpotential after the SUSY breaking, as a
soft breaking parameters (in the minimal N=1 Supergravity version
they are the universal gravitino mass $m_{3/2}$).

The superootential (5) has a minimum:  \begin{eqnarray} < \Sigma > =
diag\biggl [ ~1~ 1~ 1~ -1~ -1~ -1~ 0 ~0 ~0 \biggr ] \cdot
\frac{M_H}{h} + diag\biggl [ ~0 ~0 ~0 ~0 ~0 ~1 ~-1 ~0 \biggr ] \cdot
\frac{M_{\Phi}}{a} , \nonumber  \\ \nonumber
              \\ <H>=\left[ \begin{array}{ccccccc} 0 & 0 & 0 & 0 & 0
& 0 & 0 \\ 0 & 0 & 0 & 0 & 0 & 0 & 0 \\ 0 & 0 & 0 & 0 & 0 & 0 & 0 \\
0 & 0 & 0 & 0 & 0 & 0 & 0 \\ 0 & 0 & 0 & 0 & 0 & 0 & 0 \\ 0 & 0 & 0 &
0 & 0 & V & 0 \\ 0 & 0 & 0 & 0 & -V & 0 & 0 \\ 0 & 0 & 0 & 0 & 0 & 0
& 0 \end{array} \right] ,~~<\overline {H}>=\left[
\begin{array}{ccccccc} 0 & 0 & 0 & 0 & 0 & 0 & 0 \\ 0 & 0 & 0 & 0 & 0
& 0 & 0 \\ 0 & 0 & 0 & 0 & 0 & 0 & 0 \\ 0 & 0 & 0 & 0 & 0 & 0 & 0 \\
0 & 0 & 0 & 0 & 0 & 0 & 0 \\ 0 & 0 & 0 & 0 & 0 & -V & 0 \\ 0 & 0 & 0
& 0 & V & 0 & 0 \\ 0 & 0 & 0 & 0 & 0 & 0 & 0 \end{array} \right] ,
\nonumber \\ \nonumber \\ <\Phi _1>=\left[ \begin{array}{c} 0 \\ 0 \\
0 \\ 0 \\ 0 \\ 0 \\ V_1 \\ \end{array} \right] ,~~<\overline
{\Phi}_1>=\left[ \begin{array}{cccccccc} 0 & 0 & 0 & 0 & 0 & 0 & V_1
& 0 \end{array} \right] , \nonumber \\ \nonumber \\ <\Phi _2>=\left[
\begin{array}{c} 0 \\ 0 \\ 0 \\ 0 \\ 0 \\ 0 \\ 0 \\ V_2 \end{array}
\right] ,~~<\overline {\Phi}_2>=\left[ \begin{array}{cccccccc} 0 & 0
& 0 & 0 & 0 & 0 & 0 & V_2 \end{array} \right] ,\nonumber \\ \nonumber
\\ where~ V=\biggl [\frac{\sigma}{ah}M_{\Phi}\biggl
(\frac{4M_H}{h}-\frac{M_{\Phi}}{a}\biggr )\biggr ]^{1/2}
,~~V_1=\biggl [\frac{4\sigma}{ah}M_H
\biggl(\frac{M_H}{h}-\frac{M_{\Phi}}{a}\biggr )\biggr ]^{1/2}
,~~V_2=\sqrt{\frac{4\sigma}{bh^2}}M_H \end{eqnarray}

As in the case considered in sec.2 one pair of doublets from
$H+\overline{H}$, namely $H_{[w8]}+\overline{H}^{[w8]}$ (where
w=4,5) again remain light (in order to ignore the small masses of
$\Sigma ^2$ and $\Phi _2\overline{\Phi }_2$ exactly massless). Mass
parameter $M_\Phi $ does not play any role in the mass formation of
doublets $H_{[w8]}+\overline{H}^{[w8]}$ and could be as large as we
need. In the case $M_\Phi >M_H$ the hierarchy of SU(8) breaking is
the following:  \begin{equation} SU(8)\stackrel{\sim
M_\Phi}{\longrightarrow }SU(6)\stackrel{\sim (M_\Phi
M_H)^{1/2}}{\longrightarrow}SU(5)\stackrel{\sim
M_H}{\longrightarrow}G_{SM} \end{equation}
\begin{figure}[t]
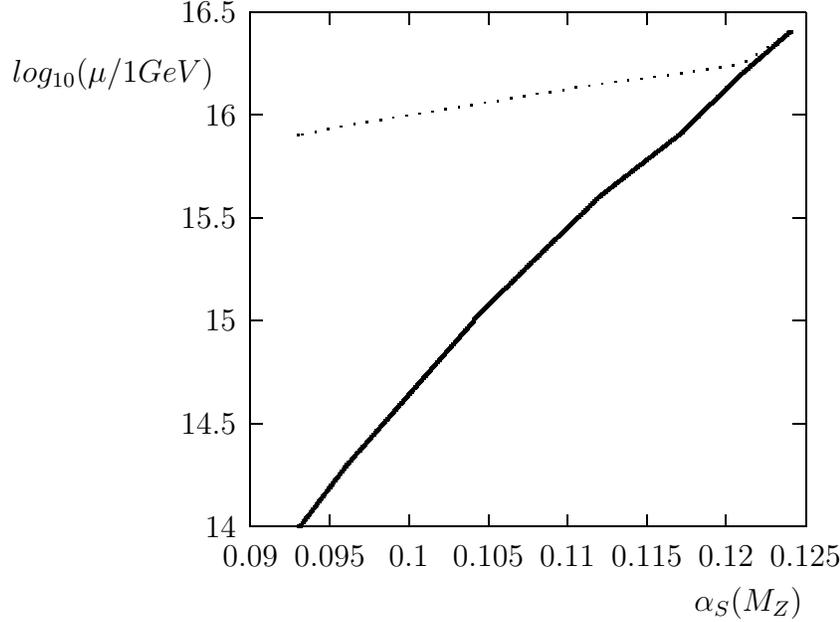

\begin{center}
\setlength{\unitlength}{0.240900pt}
\ifx\plotpoint\undefined\newsavebox{\plotpoint}\fi
\sbox{\plotpoint}{\rule[-0.175pt]{0.350pt}{0.350pt}}%

\end{center}
\caption{\small {The value of $\alpha _S(M_Z)$ predicted from the
gauge coupling unification requirement at two-loop level as a
function of $M_\Phi$ (solid line) and $M_G$ (dotted line). The
following parameters are used as input: $M_{SUSY}=M_Z, sin^2\theta
_W(M_Z)=0.2312$ and $\alpha ^{-1}_{EM}(M_Z)=127.9$} } \end{figure}
and therefore the standard MSSM unification picture up to the GUT
threshold corrections is retained. But there are solutions of RGE for
gauge $M_\Phi <M_H$ compatible with unification condition. As one can
see from Fig.2 the general tendency is the following: the decreasing
of $M_\Phi $ leads to the decreasing of the strong coupling constant
value $\alpha _S(M_Z)$, so the model in principle could predict the
value of strong gauge coupling in agreement with those extracted from
low-energy experiments.

The "custodial symmetry" mechanism demonstrated here explicitly on
the SU(8) example can be implemented for any $SU(N\geq 8)$ SUSY GUTs
by the appropriate choice of Higgs superfields and following the
general prescriptions discussed above.

\section{Discussion}

I have shown that it is possible to solve the doublet-triplet
splitting problem in a natural and economical way in the frames of
extended $SU(N\geq 8)$ SUSY GUTs. While concentrating the Higgs
sector of the model I said nothing about the quarks and leptons
(matter sector). As it is well known, the family or, in general,
flavour problem, which relates to the questions such as quark-lepton
family replication, hierarchy of quark-lepton masses and mixings,
suppression of the flavour changing neutral currents, side by side
with the gauge hierarchy problem is one of the most difficult to
explain. Very interesting framework for the solution of the flavour
puzzle is the family (horizontal) symmetries, namely local chiral
$SU(3)_H$ one [18]. So for flavour unified GUTs it is natural to
consider $SU(N\geq 8)$ symmetries. This offers an intriguing
possibility to solve the both gauge hierarchy and flavour problem in
the frames of extended $SU(N\geq 8)$ SUSY GUTs [19].

\section{Aknowledgements}

I would like to thank J.L.Chkareuli, I.Gogoladze and Z.Tavartkiladze
for discussions and for invitation and warm hospitality the
organizers of ICTP Summer School in High Energy Physics and Cosmology
(Trieste, Italy) where part of this work was made. I also acknowledge
the partial support of the Georgian Government and ISF under the
Grant No.MXL200.

\end{document}